\documentclass{article}

    \usepackage[preprint]{neurips_2019}

\usepackage[utf8]{inputenc} 
\usepackage[T1]{fontenc}    
\usepackage{hyperref}       
\usepackage{url}            
\usepackage{booktabs}       
\usepackage{amsfonts}       
\usepackage{nicefrac}       
\usepackage{microtype}      
\usepackage{amsmath}        
\usepackage[pdftex]{graphicx}

\title{Translating neural signals to text using a Brain-Machine Interface}

\author{%
	Janaki Sheth\\
	Department of Physics and Astronomy\\
	University of California, Los Angeles\\
	Los Angeles, CA 90095 \\
	\texttt{janaki.sheth@physics.ucla.edu} \\
	\And
	Ariel Tankus\\
	Department of Neurology and Neurosurgery \\
	Tel Aviv University\\
	Tel Aviv, Israel \\
	\texttt{arielta@post.tau.ac.il} \\
	\And
	Michelle Tran\\
	Department of Neurosurgery \\
	University of California, Los Angeles\\
	Los Angeles, CA 90095 \\
	\texttt{metran@mednet.ucla.edu} \\
	\And
	Nader Pouratian\\
	Department of Neurosurgery \\
	University of California, Los Angeles\\
	Los Angeles, CA 90095 \\
	\texttt{npouratian@mednet.ucla.edu} \\
	\And
	Itzhak Fried\\
	Department of Neurosurgery \\
	University of California, Los Angeles\\
	Los Angeles, CA 90095 \\
	\texttt{ifried@mednet.ucla.edu} \\
	\And
	William Speier\\
	Department of Radiology \\
	University of California, Los Angeles\\
	Los Angeles, CA 90095 \\
	\texttt{speier@ucla.edu} \\
}

\begin{document}
	
	\maketitle
	
	\begin{abstract}
		Brain-Computer Interfaces (BCI) help patients with faltering communication abilities due to neurodegenerative diseases produce text or speech output by direct neural processing. However, practical implementation of such a system has proven difficult due to limitations in speed, accuracy, and generalizability of the existing interfaces. To this end, we aim to create a BCI system that decodes text directly from neural signals. We implement a framework that initially isolates frequency bands in the input signal encapsulating differential information regarding production of various phonemic classes. These bands then form a feature set that feeds into an LSTM which discerns at each time point probability distributions across all phonemes uttered by a subject. Finally, these probabilities are fed into a particle filtering algorithm which incorporates prior knowledge of the English language to output text corresponding to the decoded word. Performance of this model on data obtained from six patients shows encouragingly high levels of accuracy at speeds and bit rates significantly higher than existing BCI communication systems. Further, in producing an output, our network abstains from constraining the reconstructed word to be from a given bag-of-words, unlike previous studies. The success of our proposed approach, offers promise for the employment of a BCI interface by patients in unfettered, naturalistic environments.
	\end{abstract}
	
	\section{Introduction}
	
	Neurodegenerative diseases such as amyotrophic lateral sclerosis (ALS) restrict an individual's potential to fully engage with his or her surroundings by hindering their communication abilities. Brain-Computer Interfaces (BCI) have long been envisioned to assist such patients as they bypass the affected pathways and directly translate neural recordings into text or speech output \citep{Brumberg18}. These devices are trained to generate appropriate models of a subject's brain, and then classify and translate neural signals into commands. 
	
	However, practical implementation of this technology has been hindered by limitations in speed and accuracy of existing systems \citep{Farwell88}. Many patients rely on communication devices that use motor imagery \citep{Mcfarland00}, or on interfaces that require them to individually identify and spell out text characters such as the "point and click" cursor method \citep{ Pandarinath17,Speier17,Townsend16}. Despite significant work in optimizing these systems, the inherent limitations in their designs restrict them to communication rates far less than naturalistic speech \citep{Huggins11}. 
	
	To address these shortcomings, several studies are using electrocorticography (ECoG) and local field potential (LFP) signals for classification and reconstruction of individual phonemes and their acoustic features \citep{Akbari19,Pasley12}. These invasive approaches provide superior signal quality as neural recordings are taken directly on top of the cortex or within the cortical layer, thus capturing events from indivdual cells or small populations of neurons with high temporal and spatial accuracy. Previous work attempted translation to continuous phoneme sequences using invasive neural data \citep{Herff15,Moses16}; however, despite their reported higher translation speed, their applications are limited to a reduced dictionary (10-100 words). Other design choices meant to enhance phoneme classification capitalize on prior knowledge of the target words, hindering their generalization to unmodified, naturalistic scenarios. Additionally, a recent study synthesized speech directly using recordings from the speech cortex. Though it demonstrates partial transferrability of its decoder amongst patients, the accuracy of said model is again limited to selection of the reconstructed word by a listener from a given pool of 25 words and worsens as the pool size increases \citep{Anumanchipalli19}. 
	
	Thus, establishing the capability of these approaches to generalize to unconstrained vocabularies is not obvious and has to our knowledge not yet been studied. Here, we present the performance of a two-part decoder network comprising of an LSTM and a particle filtering algorithm on data gathered from six patients. We provide empirical evidence that our algorithm achieves an average accuracy of 32\% using a generalized language model based on the expansive Brown corpus, marking an important, non-incremental step in the direction of viability of this interface. 
	
	\section{Methods}
	
\begin{figure}
		\centering
		\includegraphics[width = 0.75\linewidth]{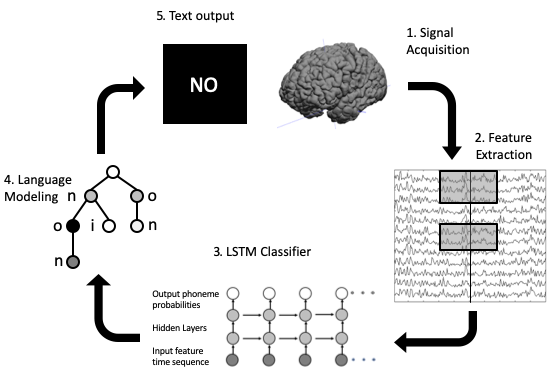}
		\caption{Overview of the process for translating neural signals into text. 1) Signals are recorded using depth electrodes implanted based on clinical need. 2) Signal features are selected for each time point based on spectral analysis. 3) A bi-directional LSTM is used to create probability distributions over phonemes at each time point. 4) Probabilities are smoothed and domain knowledge is incorporated using a language model. 5) The highest probability word is chosen as the output.}
		\label{fig:overview}
	\end{figure}

	The overall system for translating neural signals into text consists of five steps (Fig.\ref{fig:overview}). First, signals are recorded from depth electrodes implanted for surgical treatment of epilepsy while patients were instructed to speak individual words. Then, spectral features are extracted from these signals to create feature vectors for classification. An LSTM classifier uses these to generate probability distributions over the set of phonemes at each time point. Then, a partical filtering algorithm temporally smooths the probabilities incorporating prior knowledge from a language model. Finally, the output text is produced and compared with the original spoken words. Each step is further elaborated upon in the following sections.
	
	\subsection{Experimental Design}
	Data was obtained from neurosurgical patients implanted with intracranial depth electrodes to identify seizure foci for potential surgical treatment of epilepsy \cite{Tankus12}. Implantation of electrodes in the relevant areas of the temporal, frontal, and parietal lobes was based on clinical need. This study was approved by the institutional review board and all subjects consented to participate in this research.
	
	During the study, subjects were asked to repeat individual words ("yes", "no"), or singular vowels with or without preceding consonants. During each trial, they were told which word or string to repeat. Then, they would be prompted by a beep followed by a 2.25 second window during which they repeated the word. The number of trials varied between subjects based on their comfort, resulting in a variable number of trails ranging from 55 to 208. Consequently, the number of phonemes per subject varied from 8 (3 consonants, 5 vowels) to 10 (5 consonants, 5 vowels). The sampling rate of these recordings was 30 kHz. Before further processing, electrodes determined visually to have low signal-to-noise ratio (SNR) were removed.  
	
	\subsection{Feature Selection}
	
\begin{figure}
		\centering
		\includegraphics[width = 0.8\linewidth]{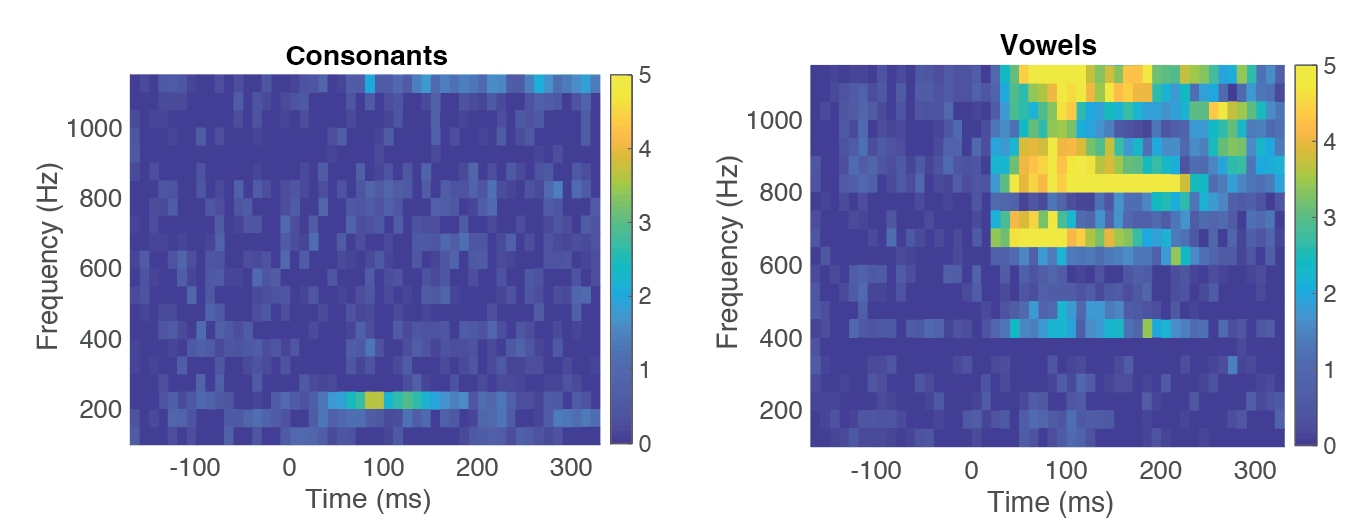}
		\caption{The time frequency maps for subject 4 illustrate bands tuned for consonants and vowels respectively, represented using z-scores. 0 ms corresponds to the start of speech.}
		\label{fig:frequency bands}
	\end{figure}

	Since we sought to include as input to our decoder differential information stored in the ECoG signals about production of various phonemes, we designed an experiment that mapped power in frequency bands of the neural recordings to the underlying phoneme pronunciation. The motivation for this experiment was that previous studies \citep{Chang13,Herff15,Moses16} have used bands up to high gamma (70-150 Hz) to map unto underlying speech, but our preliminary analysis found tuning in a greater range of frequencies for several of our subjects (Fig.~\ref{fig:frequency bands}).
	
	Each recording was divided into time windows from -166.67 to 100 ms relative to onset of the speech stimuli. Labels [0,1] were assigned respectively to the corresponding audio signal: [silence, consonant/vowel]. The power per band is pre-processed by z-scoring and then down sampled to 100 Hz. This further acts as an input to a linear classifier which we train using early-stopping and coordinate descent methods. To additionally ensure that the classifier can correctly identify the silence after completion of the phoneme string, we performed training over 100 ms post speech onset, but test the features captured by the classifier weights over 333.33 ms, since most trials end within this time period. 
	
	The input feature set for each subject thus is a concatenated matrix comprising of the time domain signal, a frequency band with the highest z-score for vowel production and lastly, a band encapsulating consonant information. The requisite frequency bands are: Subjects 1 - [150-200, 200- 250], 2 - [150-200, 1000-1150], 3 - [70-150], 4- [200-250, 650-1150], 5- [200-250, 700-1150], 6 - [150-400, 600-750]. This algorithm was implemented using the STRF toolbox \citep{STRF} in Matlab \citep{Matlab}.
	
	\subsection{LSTM Model Description}
	
	The first part of our decoder is a stacked two-layer bLSTM which takes as input the feature set and outputs a probability distribution across all phonemes in the given dataset. We use a bLSTM due to its ability to retain temporally distant dependencies when decoding a sequence \cite{graves05}. Further, our analysis reveals that while a single-layer network can differentiate between phonemic classes such as nasals, semivowels, fricatives, front vowels and back vowels; a two-layer model can distinguish between individual phonemes. There are 256 hidden units for each LSTM cell. The model is trained using the ADAM optimizer to minimize weighted cross-entropy error, with weights inversely proportional to the phoneme frequencies. The optimizer is initialized with beta1 = 0.9, beta2 = 0.999 and eps = $1e^{-8}$. Training occurs over 40 epochs with a learning rate of $1e^{-3}$. Using leave-one-out cross validation, the recurrent network outputs a time sequence of the probability distributions. Software was implmented using Pytorch.

	\subsection{Language Model}
	A language model is used to apply prior knowledge about the expected output given the target domain of natural language. In general, such a model creates prior probability distributions for the output based on the sequences seen in a corpus that reflects the target output. In this study, word frequencies were determined using the Brown corpus, which contains over 2 million words compiled from various types of documents published in the United States in 1961 \cite{Francis79}. These words were translated into their corresponding phonemic sequences using the CMU Pronouncing Dictionary \cite{Weide98}. For words with multiple pronunciations, one of the possibile pronunciations was randomly chosen for each occurrance of the word. Phoneme prior probabilities were determined by finding the relative frequency of each phoneme in the resulting corpus. 
	
\begin{figure}
		\centering
		\includegraphics[width = 0.35\linewidth]{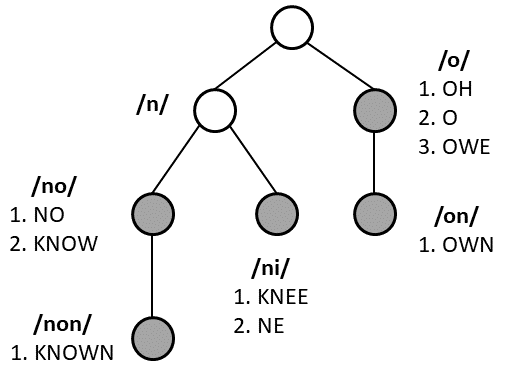}
		\caption{Example language model based on only three phonemes: /n/, /o/, and /i/. Shaded nodes correspond to complete words, which link back to the root. Lists of words correspond to homophones that a given node can represent.}
		\label{fig:language model}
	\end{figure}

	To find probabilities of sequences of phonemes, these prior probabilities can be simplified using the nth-order Markov assumption to create an n-gram model \cite{Speier11,Manning99}. While n-gram models are able to capture local phonemic patterns, they allow for sequences that are not valid words on the language. A probabilistic automaton (PA) creates a stronger prior by creating states for every subsequence that starts a word in the corpus \cite{Speier15}. Thus, the word “no” would result in three states: \textbackslash n\textbackslash,  \textbackslash no\textbackslash, and the start state which corresponds to a blank string. Each state then links to every state that represents a superstring that is one character longer. Thus, the state \textbackslash n\textbackslash,  will also link to the state \textbackslash ni\textbackslash  (Fig. \ref{fig:language model}).
	
	In the English language, it is possible for multiple words with different spellings to have identical pronunciation (homophones). The language model accounts for this possibility by keeping a list of the words associated with each node in the model along with their relative frequency in the text corpus. In the current implementation, the model will select the highest probability word associated with the selected node. While this process could lead to incorrect selections in practice if the intended target were a less common homophone, deciding between these options would require a language model that incorporates context that extends beyond single words (see future directions).
	
	\subsection{Temporal Smoothing}
	Laplacian smoothing is applied to the output of the LSTM model so that phonemes that were not seen during training are assigned a non-zero probability. A temporal model is then used to apply the language model to the resulting distributions. For simple n-gram based models, dynamic programming methods such as hidden Markov models can be implemented for this purpose. As more sophisticated language models are used, the ability to fully represent the probability distribution over possible output sequences becomes impractical.
	
	In this study, we applied a particle filtering (PF) method previously applied in P300-based brain computer interface systems \cite{Speier15}. PF is a method for estimating the probability distribution of sequential outputs by creating a set of realities (called particles) and projecting them through the model based on the observed data \cite{Gordon93}. Each of these particles contains a reference to a state in the model, a history of previous states, and an amount of time that the particle is going to remain in the current state. The distribution of states occupied by these particles represents an estimation of the true probability distribution.
	
	When the system begins, a set of P particles is generated and each is associated with the root node of the language model. At each time point, samples are drawn from the proposal distribution defined by the transition probabilities from the previous state.
	
	\begin{equation}
	x_t^{(L)}\sim p(x_t\mid x_{t-1}^{(L)})]   
	\end{equation}
	
	The time that the particle will stay in that state is drawn from a distribution representing how long the subject is expected to spend speaking a specific phoneme. At each time point, the probability weight is computed for each of the particles using,
	
	\begin{equation}
	w_t^{(L)}\propto w_{t-1}^{(L)} p(y_t\mid x_t^{(L)} )
	\end{equation}
	
	The weights are then normalized and the probability of possible output strings is found by summing the weights of all particles that correspond to that string. The system keeps a running account of the highest probability output at each time. The effective number of particles is then computed.
	
	\begin{equation}
	P_{eff}=\frac{1}{(\Sigma_i (w_t^{(L)} )^2 )}
	\end{equation}
	
	If the effective number falls below a threshold, $P_{thresh}$, a new set of particles are drawn from the particle distribution. At each time point, the amount of time for a given particle to remain in a state is decremented. Once that counter reaches zero, the particle transitions to a new state in the language model based on the model transition probabilities $p(x_t\mid x_{0:t-1} )$.

	\subsection{Evaluation}\label{Evaluation}
	The simplest evaluation metric used is the trial accuracy, which represents the number of trials classified completely correctly, divided by the total number of trials. Trials are only considered correct if the phoneme sequence matches the labels and each of those phonemes overlaps at least partially with the corresponding label.
	
	Phoneme-based performance was measured in terms of precision, recall, and phoneme error rate. Here, each phoneme classification was considered either a true positive (i.e., correct phoneme overlapping the label), false positive (i.e., classified phoneme that either doesn’t match the corresponding label or occurs during silence), or false negative (i.e., no detected phoneme during a label). Precision is then the number of true positives divided by the sum of the true positives and false positives. Recall is the number of true positives divided by the sum of the true positives and false negatives. Phoneme error rate is the number of changes that would need to be made to the output sequence in order to match the label sequence (also known as the Levenshtein distance) divided by the length of the label sequence \cite{Moses16}.
	
	For evaluation of the output as a BCI system, we must take into account two factors: the ability of the system to achieve the desired result and the amount of time required to reach that result. Because there is a trade-off between speed and accuracy, evaluation in BCI communication literature is traditionally based on the mutual information between the selected character, x, and the target character, z, referred to as the bits per symbol (B). 
	
	\begin{equation}
	B=\Sigma_z p(z) \Sigma_x p(x\mid z) \log \frac{p(x\mid z)}{p(x)}
	\end{equation}	
	
	In the most common metric, information transfer rate (ITR), the probabilities for all characters are assumed to be the same (p(x)=1/N where N is the size of the alphabet) and errors are assumed to be uniform across all possible characters, reducing the bits per symbol to
	
	\begin{equation}
	B_C = \log N + ACC_C + (1 - ACC_C)\log \frac{1-ACC_C}{N-1}
	\end{equation}
	
	Where $ACC_C$ is the accuracy of individual character selections. Thus the ITR given the average number of characters selected per minute ($CPM = \frac{n}{time}$) is $ITR=B_C*CPM$ \cite{pierce80}.
	
	It has previously been observed that ITR overestimates the amount of information conveyed by the system because characters do not occur with equal frequency \cite{speier13}. Also, the amount of information that ITR assigns to a word is based largely on the word’s length. This metric assigns a significantly higher amount of information to incorrect strings that share characters with the target, regardless of whether they make syntactic sense or possibly confuse the meaning. An alternative would be to base the metric on word frequency $(p(z)= \frac{c(z)}{c(*)})$. The accuracy can then be computed as the fraction of correct words ($ACC_W= \frac{\Sigma_t \delta_{x_t}^{z_t } }{n}$), resulting in a conditional probability of a selection. The bits per symbol ($B_W$) then becomes
	
	\begin{equation}
	B_W=\Sigma_z p(z)(ACC_W  \log \frac{ACC_W}{p(z)} + (1-ACC_W )  \log\frac{1-ACC_W}{1-p(z)})
	\end{equation}
	
	Multiplying this by words selected per minute ($WPM=\frac{n}{time}$) gives a bit rate based on mutual information (MI). 
	
	\begin{equation}
	MI=B_W*WPM
	\end{equation}
	
	Because the distributions for speeds, accuracies, and bit rates are not normally distributed, significance was tested for all metrics using Wilcoxon signed-rank tests.
	
	\section{Results}
	
	Word accuracies varied between subjects, ranging from $54.55\%$ (subject 1) to $13.46\%$ (subject 2) (Table \ref{Table1}). On average, $32.16\%$ of trials were classified completely correctly and an additional $23.06\%$ had at least one phoneme match. Of the incorrect classifications $32.28\%$ produced incorrect words either because none of the output phonemes were correct or because the sequences did not align temporally with the audio signal. In the remaining $12.49\%$ of trials, the system did not detect speech signals, and produced an empty string as output.
	
	\begin{table}[!ht]
		\caption{Word level performance of each subject. }
		\label{Table1}
		\centering
		\begin{tabular}{lcccc}
			\hline
			Subject    & $ACC_W$ ($\%$)       & Partial ($\%$)        & Incorrect ($\%$) & Omission ($\%$)     \\ \hline \hline
			1  & 54.55  & 12.73  & 20.00  & 12.73   \\
			2  & 13.46  & 21.15  & 46.15  & 19.23   \\
			3  & 23.08  & 35.75  & 36.65  &  4.52   \\
			4  & 29.73  & 18.92  & 32.43  & 18.92   \\
			5  & 31.43  & 24.57  & 33.71  & 10.29   \\
			6  & 40.72  & 25.26  & 24.74  &  9.28   \\
			\hline
			Average  & 32.16  & 23.06  & 32.28  & 12.49 \\ \hline
			\hline
		\end{tabular}
	\end{table}
	
\begin{figure}
		\centering
		\includegraphics[width = 0.6\linewidth]{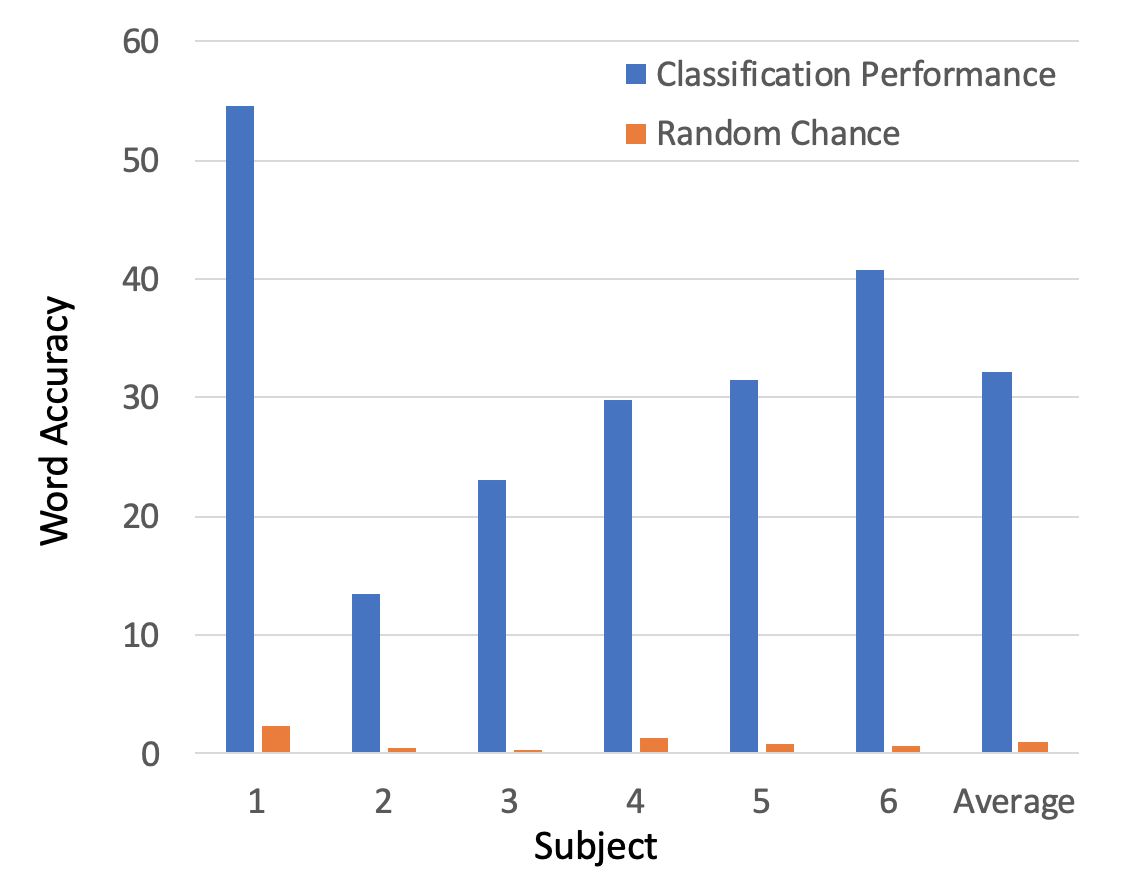}
		\caption{Word classification accuracy for each subject compared to the expected value based on random chance. Chance accuracies were determined by randomly selecting words based on probabilities provided by the language model.}
		\label{fig:word accuracy}
	\end{figure}

	Expected word accuracy for each subject was computed by finding the expected value of having the output match the target word given only the language model. Though for this computation, the language model was simplified to include only those words from the Brown corpus that could be possibly constructed using the phonemes that a subject uttered, this is a reasonable restraint. Additionally, it is far more lenient as compared to previous studies wherein the output word is constrainted to only a small subset of their analogously feasible word pool. Comparing each of the word accuracies to the expected results from random signals, we found that all subjects perform significantly better than random (p<0.01) (Figure \ref{fig:word accuracy}).
	
	On average phoneme classification yielded precision, recall, and error rates of 0.46, 0.51, and 73.32, respectively (Table~\ref{Table2}). The higher recall suggests that more errors were a result of incorrectly adding phonemes to an output sequence than missing phonemes in the classification. The phoneme error rates that we observed in our output sequences were lower on average than those reported previously by Moses et al.~\cite{Moses16}.
	
	\begin{table}[!ht]
		\caption{Phoneme level performance of each subject. }
		\label{Table2}
		\centering
		\begin{tabular}{lccc}
			\hline
			Subject    & Precision       & Recall        & Phoneme Error Rate     \\ \hline \hline
			1  & 0.63  & 0.76  & 41.25   \\
			2  & 0.27  & 0.31  & 94.70   \\
			3  & 0.40  & 0.53  & 87.94   \\
			4  & 0.45  & 0.41  & 80.63   \\
			5  & 0.45  & 0.52  & 74.80   \\
			6  & 0.58  & 0.52  & 60.58   \\
			\hline
			Average  & 0.46  & 0.51  & 73.32 \\
			\hline
			Moses et al. \cite{Moses16} & - & - & 87.56 \\
			\hline \hline
		\end{tabular}
	\end{table}
	
	
	To compare the performance of this system with that achieved by existing ERP-based BCI systems, we calculated the bit rate for communication using the mutual information metric described in section \ref{Evaluation}. For the word selection rate, we used the full time that a subject was given to speak a word (2.25 seconds), resulting in a WPM value of 26.67 words per minute for each subject. Word accuracies varied between subject, ranging from $13.46\%$ to $54.55\%$, with an average value of $32.15\%$ resulting in an average MI of $47.35$ bits per minute. This value was significantly higher than the results presented in \cite{Speier17}. Since the study presented by Townsend et al \cite{Townsend16} used several different configurations for their subjects, we compare our results here with the best performing subject in their study. The average MI value in this study was over twice the value achieved by their best subject, with all but one subject in this study achieving a higher bit rate.
	
	\begin{table}[!ht]
		\caption{BCI communication performance in terms of words per minute ($WPM$), word accuracy ($ACC_W$), bits per word ($B_W$), and mutual information ($MI$). }
		\label{Table3}
		\centering
		\begin{tabular}{lcccc}
			\hline
			Subject    & $WPM$       & $ACC_w$ ($\%$)        & $B_W$ & $MI$     \\ \hline \hline
			1  & 26.67  & 54.55  & 3.31  & 88.36  \\
			2  & 26.67  & 13.46  & 0.60  & 16.01  \\
			3  & 26.67  & 23.08  & 1.16  & 30.91  \\
			4  & 26.67  & 29.73  & 1.58  & 42.07  \\
			5  & 26.67  & 31.43  & 1.69  & 45.02  \\
			6  & 26.67  & 40.72  & 2.32  & 61.73  \\
			\hline
			Average  & 26.67  & 32.16  & 1.78  & 47.35  \\ \hline
			Speier et al. \cite{Speier17}  & 2.53  & 92.56  & 6.54  & 16.54  \\ \hline
			Townsend et al.* \cite{Townsend16}  & 2.94  & 100.00  & 7.33  & 21.56  \\ \hline
			\hline
		\end{tabular}
		
		*performance of best single subject in study
	\end{table}
	
	\section{Discussion}
	
	Each of the subjects in this study were able to communicate with significantly higher accuracy than chance. Nevertheless, the average word error rate seen in this study (67.8\% on average) was higher than the 53\% reported in \cite{Anumanchipalli19}. There were several important differences in these studies, however. The primary difference is that their system produced an audio output that required a human listener to transcribe into a word selection. Despite advances in machine learning and natural language processing, humans have superior ability to use contextual information to find meaning in a signal. Furthermore, that study limited classifications to an output domain set of 50 words, which is generally not sufficient for a realistic communication system. 
	
	The communication speeds reported here are based on the trial time of 2.25 seconds. This time was set conservatively to make sure that subjects had time to respond to prompts, and the majority of the time was spent waiting for the next speaking cue. The actual time spent speaking the promted words was under 400 ms on average across subjects. This speaking time is in line with the average rate of natural speech, which is usually reported to be in the range of 100-125 words per minute \cite{kemper94}. Increasing the rate of word production would further improve the bit rate of a speech decoding system in comparison to existing BCI spellers.
	
	While this study showed significant improvements over existing BCI systems in terms of bit rate and speed, our accuracies are lower than those reported in ERP-based BCI studies \cite{Townsend16,Speier17}. It has been previously reported that BCI users expect an accuracy level that exceeds $90\%$ \cite{Huggins11}, which is higher than the accuracy values achieved here. In order for a BCI system based on translating neural signals to become a practical BCI solution, improvements need to be made either in signal acquisition, machine learning translation, or user strategy. One approach could be to sacrifice some of the speed advantages by having users repeat words multiple times. While this would reduce communication speeds below natural speeking rates, it would still greatly exceed ERP-based methods, while increasing the signals available for classification which could improve system accuracy.
	
	\subsection{Limitations and Future Work}
	The language model used in this study was designed to be general enough for application in a realistic BCI system. This generality may have been detrimental to the performance in the current study, however. Language models are designed to introduce bias into a system based on the expected output given prior knowledge. Thus, language models based on natural language will bias towards words that are common in everyday speech. The current study design, however, produced many words that are infrequent in the training corpus. For instance, the single phoneme /u/ maps to the word "ooh", which occurred only once in the full corpus. As a result, the language model actually biased away from this output, making it almost impossible to correctly classify. While it would be possible to retrain the language model on the known output words, the results would then depend on knowing the set of target words, which is not realistic for a general communication system. 
	
	The results presented in this study are promising, but they represent offline performance which does not include several factors that occur in an online implementation. For instance, offline systems do not include user feedback, which can provide additional motivation or allow the user to adjust their strategy. Also, the current study was limited to epilepsy patients, rather than the target population of ALS patients. While it would be impractical to implant electrodes for such a BCI study in the target population, testing whether the results seen in such invasive studies translate to ALS patients remains to be studied. 
	
	\section{Conclusion}
	The proposed system serves as a step in the direction of a generalized BCI system that can directly translate neural signals into written text. The system achieved bit rates that were significantly higher than the current state of the art in BCI communication. However, communication accuracies are currently insufficient for a practical BCI device, so future work must focus on improving these and developing an interface to present feedback to users.
	
	\small
	\bibliography{bibliography_nips}
	\bibliographystyle{abbrvnat}

\end{document}